\documentclass[Physsubmission, Phys]{SciPost}

\hypersetup{
    colorlinks,
    linkcolor={red!50!black},
    citecolor={blue!50!black},
    urlcolor={blue!80!black}
}

\usepackage[bitstream-charter]{mathdesign}
\DeclareSymbolFont{usualmathcal}{OMS}{cmsy}{m}{n}
\DeclareSymbolFontAlphabet{\mathcal}{usualmathcal}

\usepackage{epsfig}

\newcommand{\postscript}[2]{\setlength{\epsfxsize}{#2\hsize}
   \centerline{\epsfbox{#1}}}

\begin{document}

\begin{center}{\Large \textbf{Looking forward to forward physics at the CERN's LHC
\\
}}\end{center}

\begin{center}
Luis A. Anchordoqui\textsuperscript{1$\star$}
\end{center}

\begin{center}
{\bf 1} Department of Physics and Astronomy,  Lehman College, CUNY, NY 10468, USA\\
* luis.anchordoqui@gmail.com
\end{center}

\begin{center}
\today
\end{center}


\definecolor{palegray}{gray}{0.95}
\begin{center}
\colorbox{palegray}{
  \begin{tabular}{rr}
  \begin{minipage}{0.1\textwidth}
    \includegraphics[width=30mm]{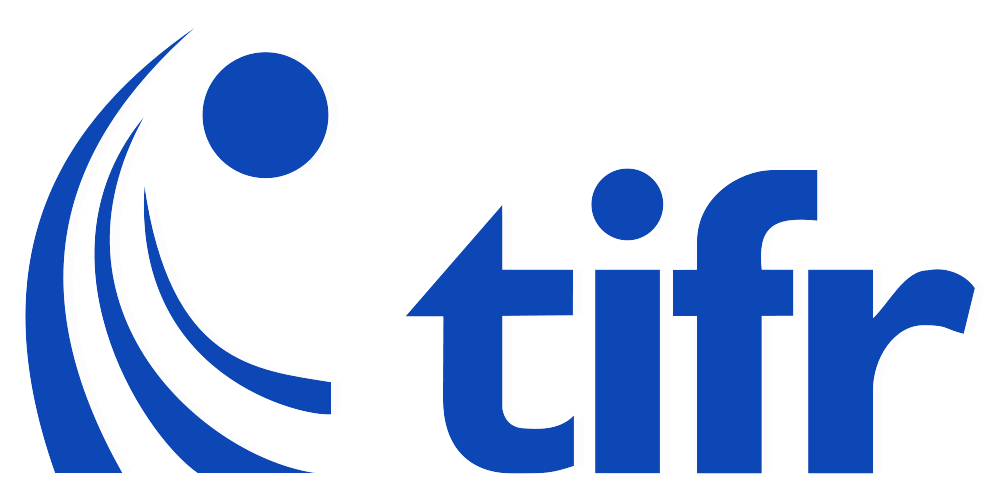}
  \end{minipage}
  &
  \begin{minipage}{0.85\textwidth}
    \begin{center}
    {\it 21st International Symposium on Very High Energy Cosmic Ray Interactions (ISVHECRI 2022)}\\
    {\it Online, 23-27 May 2022} \\
    \doi{10.21468/SciPostPhysProc.?}\\
    \end{center}
  \end{minipage}
\end{tabular}
}
\end{center}

\section*{Abstract}
{\bf For decades, new physics searches in collider
  experiments have focused on the high-$p_T$ region. However, it has
  recently become evident that the LHC physics potential has not been
  fully exploited. To be specific, forward collisions, which produce
  particles along the beamline with enormous rates, have been almost
  completely ignored. For all practical purposes, these collisions are
  a treasure trove of physics, containing the highest-energy neutrinos
  ever produced by humans, as well as possible evidence for dark
  matter, light and weakly-coupled particles, and new forces. In the
  upcoming LHC Run 3 the ForwArd Search ExpeRiment (FASER) and its
  cousin FASER$\nu$ will extend the LHC's physics potential. A
  continuation of this forward physics program for the HL-LHC aims at
  the Forward Physics Facility (FPF), with larger scale
  experiments. In this report, I give an overview of the physics
  motivations for FASER, FASER$\nu$, and the FPF experiments, including both Standard Model and beyond Standard Model physics.}

\vspace{10pt}
\noindent\rule{\textwidth}{1pt}
\tableofcontents\thispagestyle{fancy}
\noindent\rule{\textwidth}{1pt}
\vspace{10pt}

\section{Introduction}
\label{sec:intro}

For decades, new physics searches in particle colliders have focused
on the high-$p_T$ ($\gtrsim$ a few GeV) region. This is appropriate for heavy, strongly
interacting particles with small cross-sections,
${\cal O} ({\rm fb}) \lesssim \sigma_{{\rm high}-p_T} \lesssim {\cal
  O}({\rm pb})$. The Large Hadron Collider (LHC) after Run 3 will
reach a total integrated luminosity of over $300~{\rm fb}^{-1}$ and
during its high-luminosity era (HL-LHC) will reach $3~{\rm
  ab}^{-1}$. Given these considerations, a total of
$10^3 \lesssim N^{\rm events}_{{\rm high}-p_T} \lesssim 10^6$ would be
produced roughly isotropically at the HL-LHC. In the last few years it has become evident that the
LHC physics potential has not been fully exploited. Sure enough,
low-$p_T$ ($\sim \Lambda_{\rm QCD}$) forward collisions, with a total cross-section
$\sigma_{{\rm low}-p_T} \sim {\cal O} (100~{\rm mb})$, have been
almost completely ignored. For HL-LHC, this cross-section corresponds to
$N^{\rm events}_{{\rm low}-p_T} \sim 10^{17}$. The Forward Physics Facility (FPF) is one of the proposals to design and install
dedicated far-forward physics experiments at the HL-LHC~\cite{Anchordoqui:2021ghd,Feng:2022inv}. In this report, I 
give an overview of the physics motivations for the FPF experiments,
including both Standard Model (SM) and beyond Standard Model (BSM)
physics.

\section{BSM Physics Landscape}

\begin{figure}
  \postscript{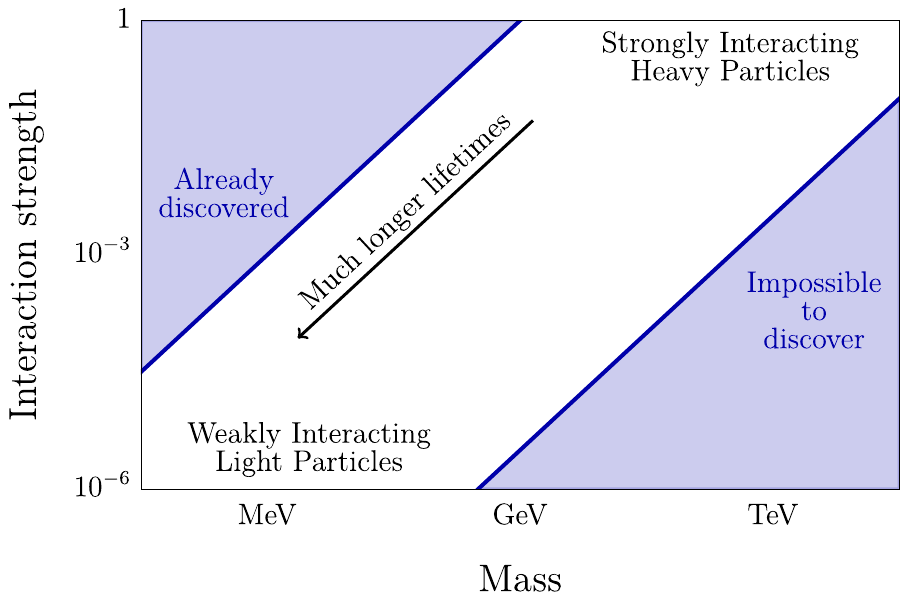}{0.5}
  \caption{The new particle landscape. Adapted
    from~\cite{Jonathan:2019}. \label{fig:1}}
  \end{figure}

The ATLAS and CMS detectors are designed to find new
heavy particles which are produced at rest and decay isotropically. However,
as more and more high-energy data are being collected and analyzed at
the LHC, the high-mass reach becomes saturated. Our best chance to
discover BSM physics is to look in places that have not been
explored before. As can be seen in
Fig.~\ref{fig:1}, this means low masses and weak
couplings. Now, these new light particles are mainly produced along
the beam line and so they would dissapear through the holes that let the beams in. We need a detector to cover
the blind spots in the forward region. High masses and weak couplings are also compelling,
but are unlikely to be accesible in the near future.

\begin{figure}
  \postscript{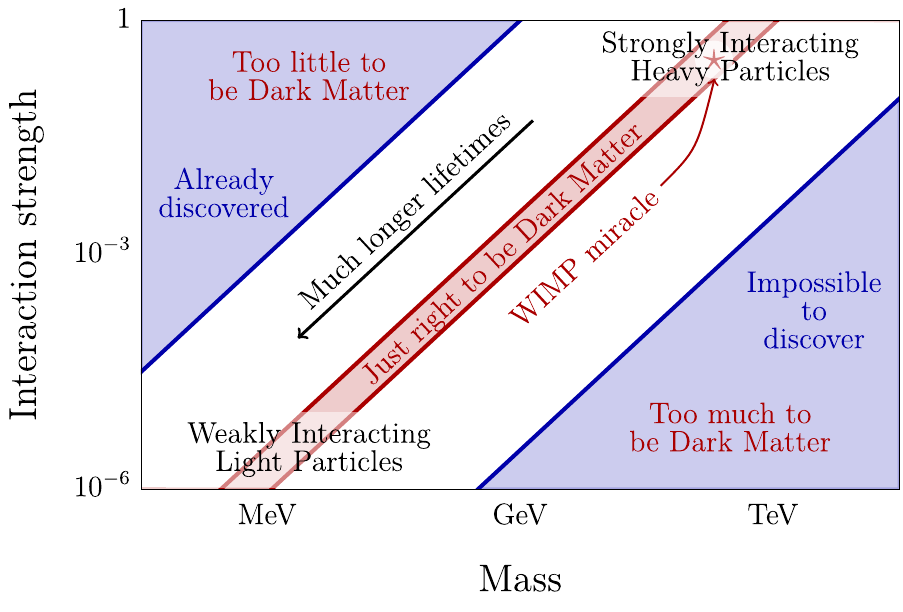}{0.5}
  \caption{The thermal relic landscape.   Adapted
    from~\cite{Jonathan:2019}. \label{fig:2}}
  \end{figure}

We have pencilled out the purely particle physics approach to new
physics searches. We can also adopt the cosmological perspective for
new physics searches and assume there is a
hidden sector with some dark matter
(DM) in it. We further assume that
the DM interacts with the SM sector through a portal, which can
be characterized by a dark photon of mass $m_{A^\prime}$ and coupling
$\epsilon$. If this were the case, then we could calculate the
interaction annihilation cross-section, which parametrically goes as
$\epsilon^2$ and by dimensional analysis scales as $m_{\rm A'}^{-2}$, yielding
\begin{equation}
  \langle \sigma v\rangle \sim \epsilon^2/m_{A^\prime}^2 \, .
  \label{sigmav}
\end{equation}  
Finally, armed with~(\ref{sigmav}) we would estimate where in the
(mass, interaction strength) plane we have the right amount of DM 
to accommodate the thermal relic density. This is shown in Fig.~\ref{fig:2} as
a red band. We know that this band coincides with the frontier
of High Energy Physics experiments, because it spans the parameter space of
the well-known {\it WIMP miracle}~\cite{Feng:2010gw}. The
parametrization given in (\ref{sigmav}) predicts that new particles
with low masses and weak
couplings inside the red band can also
accomodate the thermal relic density. It is just an incredible coincidence that the
cosmological motivation and the particle physics motivation line up so
beautifully in the (mass, interaction strength) plane.  

The dark photons may decay to visible particles but only after
traveling a long distance. The particle's velocity is near the speed
of light $v \sim 1$, its rest lifetime is enhanced by the low mass
and the small coupling $\tau \propto \epsilon^{-2} m_{\rm A^\prime}^{-1}$, and the
lifetime is further enhanced by time dilation $\gamma \propto
E/m_{A^\prime}$, hence the distance travelled is
\begin{equation}
  L = \gamma v \tau \sim 100~m \
  \left(\frac{10^{-5}}{\epsilon}\right)^2 \left(\frac{100~{\rm
        MeV}}{m_{A^\prime}}\right)^2 \left(\frac{E}{\rm TeV}\right) \, .
    \end{equation}
Altogether, weakly coupled particles imply long lifetimes, whereas low masses
typically makes particle production to peak in the forward
direction. Optimally, what we need is a precision spectrometer looking
into the forward region of LHC collisions. But of course, we cannot
place a reasonable size detector on the beamline near the interaction
point (IP), because it would block the proton beams. However, since
the particles we are looking for are long-lived and weakly-interacting
we can place a detector far upstream along the ``line of sight'' after
the beams curve. This is the idea behind the ForwArd Search ExpeRiment
(FASER), which has been installed 480~m downstream of the ATLAS IP in
the unused service tunnel and it started taking data during
Run 3~\cite{Feng:2017uoz}.\footnote{The acronym recalls
  another marvelous
  instrument that harnessed highly collimated
  particles and that was used to explore strange new
  worlds~\cite{Jonathan:2018}.} It is also of interest to ask
ourselves how big the detector has to be. For a
pseudorapidity $\eta \sim 9$, the opening angle is $\theta \simeq 2 \,
e^{-\eta} \sim 0.25~{\rm mrad}$. Therefore, most of the long-lived
particle (LLP)
signal would go through one sheet of paper at 480~m.

To appreciate the physics potential of FASER2 next we examine in detail a
specific model. The minimal dark photon portal
  with renormalizable couplings is generated by a hidden broken
  $U(1)'$ gauge symmetry whose field strength tensor, $F'_{\mu\nu}$,
  kinetically mixes with the SM hypercharge field strength tensor,
  $B_{\mu\nu}$, through the operator $F'_{\mu\nu}B^{\mu\nu}$~\cite{Holdom:1985ag}. After
  electroweak symmetry breaking, and with the gauge boson kinetic
  terms diagonalized, the dark photon $A'$ develops a suppressed coupling
  to the electromagnetic (EM) current, $J^{\mu}_{\rm EM}$. The
  renormalizable Lagrangian density of the model can be expressed in
  terms of the visible (SM) and dark sectors
\begin{equation}
  \mathcal{L} = \mathcal{L}_{\rm SM} + \mathcal{L}_{\rm dark} \,,
\end{equation}
with
\begin{equation}
	\label{eq:LgammaAp}
	\mathcal{L}_{\rm dark}  \supset
	- \frac{1}{4}F'_{\mu\nu}F^{\prime\mu\nu}
	+ \frac{1}{2} m^2_{A^\prime} A^{\prime \mu} A^{\prime}_{\mu}
	+ \epsilon \, e \,  A^\prime_\mu J^\mu_{\rm EM} \, ,
      \end{equation}
where the form of the $A^\prime$-DM interaction
$\mathcal{L}_{A^\prime \chi \bar \chi}$ is left unspecified, and where
$\chi$ denotes the DM particle. This model is
characterized by three a priori unknown parameters:  $m_{A^\prime}$, $\epsilon$, and the
decay branching fraction $\mathcal{B} (A^\prime \to \chi \bar \chi)$. Throughout we assume that the dark matter particles are
heavy such that the invisible dark-sector final states are not
kinematically allowed because $m_{A^\prime} < 2 m_{\chi}$. Given these
considerations, the phase space to be probed by FASER2 can be described in
the familiar $(m_{A'}, \epsilon)$ plane introduced in
Figs.~\ref{fig:1} and \ref{fig:2}.

\begin{figure}[t]
      \postscript{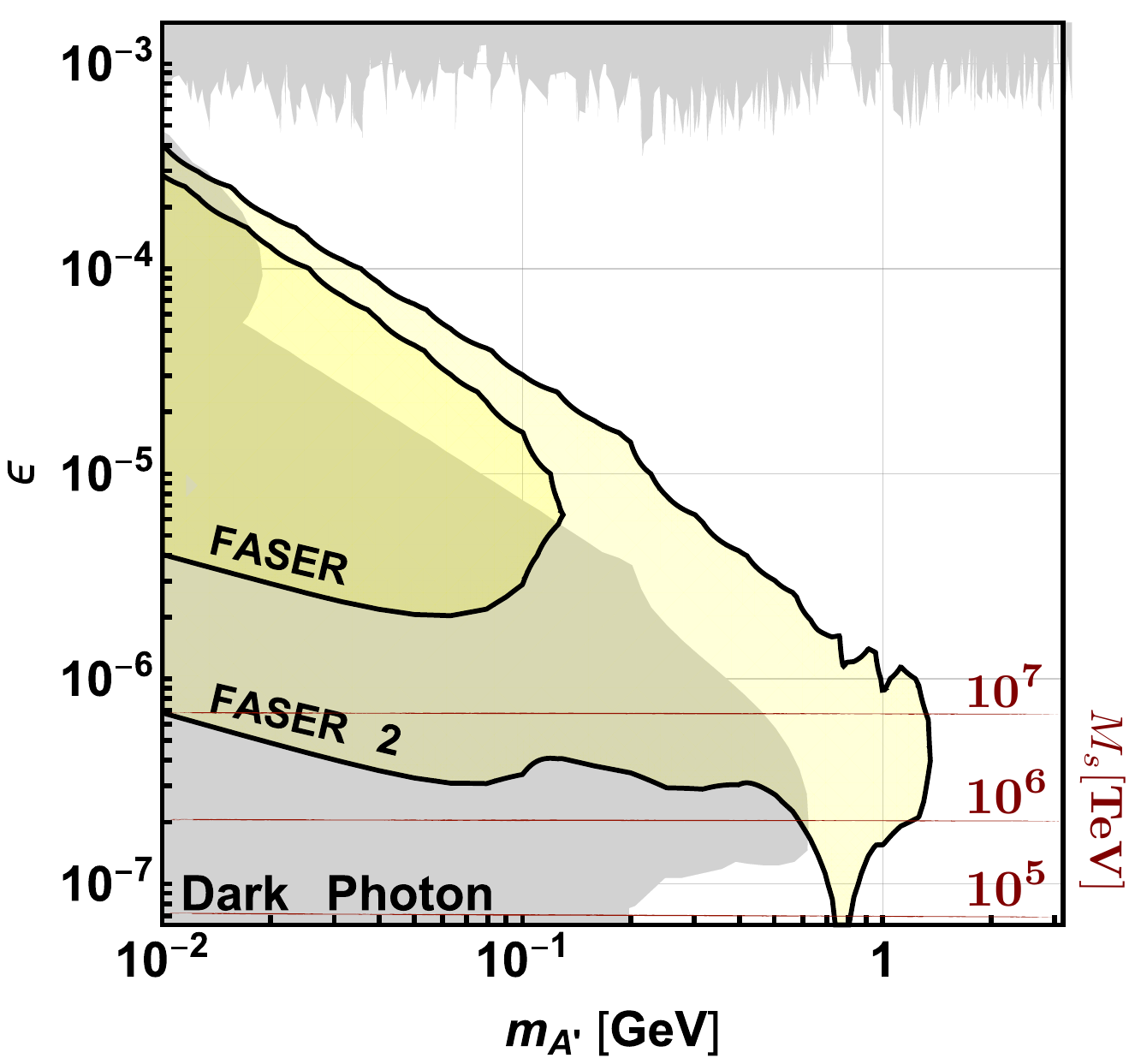}{0.5}
\caption{Dark photon sensitivity reach contour in the
        $(m_{A^\prime}, \epsilon)$ plane obtained with the {\tt FORSEE}
        package~\cite{Kling:2021fwx}. The grey shaded regions are excluded by previous
        experiments. The vertical axis on the right indicates the
        $\epsilon \leftrightharpoons M_s$ connection in LST models~\cite{Anchordoqui:2022kuw}. Adapted from~\cite{FASER:2019aik}. \label{fig:3}}
      \end{figure}

At hadron colliders like the LHC, dark photons can be abundantly produced
through proton bremsstrahlung or via the decay of heavy mesons. For
example, the expected branching fraction of $\pi^0$ decay is found to be
\begin{equation}
  {\cal B} (\pi^0 \to \gamma A^\prime) = 2 \epsilon^2 \left(1 -
    \frac{m_{A^\prime}^2}{m_{\pi^0}^2 }\right)^3 \ {\cal B} (\pi^0 \to
    \gamma\gamma) \,;
\end{equation}
similar expressions can be obtained for $\eta$, $B$, and $D$
mesons. Indeed, over the lifetime of the HL-LHC there will be $4
\times 10^{17}$ neutral pions, $6 \times 10^{16}$ $\eta$ mesons, $2
\times 10^{15}$ $D$ mesons, and $10^{13}$ $B$ mesons produced in the
direction of FASER2. The partial decay width of $A'$ to
SM leptons is given by
\begin{equation}
	\Gamma_{A'\to \ell^+\ell^-}
=	\tfrac{\epsilon^2 \alpha_{\rm EM}}{3} m_{A'} \left(  1 +2\tfrac{m^2_\ell}{m^2_{A'}} \right) \sqrt{ 1 - 4\tfrac{m^2_\ell}{m^2_{A'}} } \, ,
\end{equation}
where $\alpha_{\rm EM}$ is the fine structure constant,
$\ell=e,\mu,\tau$, and $m_{A'} > 2m_{\ell}$~\cite{Ilten:2016tkc}. The partial decay width of $A'$ to SM hadrons can be
extracted from the measured value of ${\cal R_\mu} \equiv \sigma_{e^+
  e^- \to {\rm hadrons}}/\sigma_{e^+
  e^- \to \mu^+\mu^-}$ at center-of-mass energy equal to $m_{A'}$, and
is given by 
\begin{equation}
  \Gamma_{A' \to {\rm hadrons}} = \Gamma_{A' \to \mu^+ \mu^-} \times {\cal R}_\mu (m_{A'})\, .
\end{equation}
Loop-induced SM decays ($A' \to 3\gamma$ and $A' \to \nu \bar \nu$)
are highly suppressed. In Fig.~\ref{fig:3} we show the $U(1)'$ sensitivity
reach contour of FASER2 in the $(m_{A^\prime},\epsilon)$ plane, as 
obtained with the {\tt FORSEE} package~\cite{Kling:2021fwx}.

      So far we have left the strength of the kinetic mixing
      completely unknown and model independent. We now consider a
      specific Little String Theory (LST) model, in which the
      weakness of gravity (or equivalently the enormity of the Planck
      mass) is attributed to the tiny value of the
      string coupling $g_s$~\cite{Antoniadis:2001sw}. In spite of the
      small $g_s$, in type II string theories with gauge interactions
      localized in the vicinity of Neveu-Schwarz (NS) 5-branes, the SM gauge couplings
      are of order one and are associated with the sizes of compact
      dimensions.

The
internal space is taken to be a product of a two-dimensional space, of volume
$V_2$, times a four-dimensional compact space, of volume $V_4$. The
relation between the
Planck mass $M_{\rm Pl}$ and the string scale $M_s$ is given by
\begin{equation}
 M_{\rm Pl}^2 = \frac{8}{g_s^2} \ M_s^8 \ \frac{V_2 V_4}{(2 \pi)^6} \,,
\label{mpl}
\end{equation}
with all the internal space radii of the order of the string length, i.e.,
$M_s^6 V_2 V_4 \simeq (2\pi)^6$. The hierarchy
problem is equivalent to understanding the smallness of $g_s$. The
$U(1)'$ lives on a D7-brane that wraps a 4-cycle of volume $V_4$ and
its remaining four dimensions extend into the uncompactified
space-time. We remind the reader that within LST models the SM lives
on NS5 branes, but since the $U(1)'$ gauge boson comes from D-branes the SM is generically neutral
under $U(1)'$ as desired. The $U(1)'$ gauge coupling is found to be~\cite{Anchordoqui:2022kuw}:
\begin{equation}
g^\prime \simeq \ 32^{1/4} \sqrt{\pi} \ \sqrt{M_s/M_{\rm Pl}} \, .
\label{gcoupling}
\end{equation}
The mass of the dark photon is usually generated through spontaneous
symmetry breaking; namely, through the vacuum expectation value (vev)
of a hidden SM singlet Higgs field which  carries a $U(1)'$
charge. In LST models this is possible provided
we tolerate a hierarchy between the vev and $M_s$ that
increases as $M_s^{3/2}$~\cite{Anchordoqui:2022kuw}. As noted above, the $U(1)'$ does not couple directly to the visible
sector, but does it via kinetic mixing with the hypercharge. This
coupling is generated by loops of heavy dark states carrying charges
$(q^{(i)}, q^{\prime (i)})$ under the two $U(1)$'s and having masses $M_i$,
\begin{equation}
\epsilon =\frac{e g^\prime }{16 \pi^2} \sum_i q^{(i)}
q^{\prime (i)} \ln {\frac{ M_i^2}{{\mu^2}}} \equiv \frac{e g^\prime }{16 \pi^2}
C_{\rm Log} \,,
\label{epsilon}
\end{equation}
where $\mu^2$ denotes the renormalization scale (which in string
theory is replaced by $M_s$), $e$ is the elementary charge, and where we absorbed also the constant
contribution in $C_{\rm Log} \sim
3$~\cite{Anchordoqui:2022kuw}. Substituting (\ref{gcoupling}) into
(\ref{epsilon}) we obtain the region of the parameter space to be
probed by FASER2, which is shown in Fig.~\ref{fig:3}. 

The dark sector may include more than one mediator particle coupled to
the visible sector. SM gauge and Lorentz symmetries greatly restrict the ways in which
mediators can couple to SM interactions. Portals relevant for dark sector SM interactions depend on
mediator spin and parity. FASER2 will not only be sensitive to the
spin-1 vector gauge boson, but also to
the spin-0 dark Higgs boson~\cite{Patt:2006fw}, the spin-$\frac{1}{2}$ sterile
neutrino~\cite{Pospelov:2007mp}, and pseudo-scalar axions~\cite{Nomura:2008ru}. What's more,
DM particles could scatter inside FPF dedicated neutrino experiments like FASER$\nu$2, the Advanced Scattering and Neutrino Detector (AdvSND), and
the Forward Liquid Argon Experiment (FLArE) producing visible signals~\cite{Batell:2021blf}.\footnote{The Advanced SND
  project is meant to extend the physics case of the SND@LHC
  experiment. It will consist of two detectors: one placed in the
  same $\eta$ region as SND@LHC, i.e. $7.2 < \eta < 8.4$, dubbed FAR,
  and the other one in the region $4 < \eta < 5$, dubbed NEAR. The FPF would host the FAR detector.} For example, elastic
scattering of DM particles may be visible as an excess of
neutrino-like events over the SM yield, whereas inelastic DM
scattering could appear as an excess of the ratio of
neutral-to-charged current-like events, $r = N_{\rm NC}/N_{\rm CC}$,
over the SM prediction $r \approx 0.31$. SND@LHC allows measuring $r$
with an accuracy $(\Delta r/r)_{\rm SND@LHC} = 10\%$ and 
the advanced configurations will reach
$(\Delta r/r)_{\rm AdvSND} = 1\%$. On top of the rich
experimental program to search for scattering of DM particles the FORward MicrOcharge SeArch
(FORMOSA) detector is primarily targeted at probing the kinetic mixing
portal~\cite{Goldberg:1986nk}, which allows DM from a hidden sector to become visible by
acquiring a small electric charge, often called {\it millicharge}~\cite{Foroughi-Abari:2020qar}. FORMOSA will reach sensitivity to
charges as low as $10^{-4}e$. This will allow FORMOSA to provide the
best probe of millicharged particles with mass between
$0.1 < m/{\rm GeV} < 100$. For a thorough discussion of BSM physics
searches at the FPF and a complete list
of references, see~\cite{Anchordoqui:2021ghd,Feng:2022inv}.

In closing, we note that other proposed  experiments at CERN which
could provide complementary measurements of new LLPs
include 
SHiP~\cite{Alekhin:2015byh},  FACET~\cite{Cerci:2021nlb}, and MATHUSLA~\cite{MATHUSLA:2019qpy}.

\section{Astroparticle Physics}

Experiments at the FPF will also lay out opportunities for
interdisciplinary studies at the intersection of high-energy particle
physics and astroparticle physics. This is feasible by looking at the
large flux of LHC neutrinos, which can be probed in low-background
environments at a safe distance away from the interaction point and
accelerator infrastructure. LHC neutrinos originate in the decay of
charged pions, kaons, hyperons and charmed hadrons, making the
measurement of the neutrino flux a unique probe of forward particle
production. The feasibility of such LHC neutrino measurements has
recently been demonstrated with the FASER prototype, which observed
the first neutrino interaction candidates at the
LHC~\cite{FASER:2021mtu}.

\subsection{Forward strangeness production}

Ultra-high-energy cosmic rays (UHECRs) 
supply a particle beam without financial burden to study collisions at
center-of-mass (c.m.) energies and kinematic regimes not accessible at
terrestrial colliders~\cite{Anchordoqui:2018qom}. For example, the
scattering in air of a
cosmic $^4$He nucleus with energy $E = 10^{10.3}~{\rm GeV}$ reaches a
c.m. energy in the nucleon-nucleon system of $\sqrt{s_{NN}} \sim \sqrt{2 E_N m_N} \sim 100~{\rm TeV}$, where
$E_N \simeq  10^{9.7}~{\rm GeV}$ is the cosmic ray energy per nucleon
in the fixed target frame and $m_N$ is the iso-scalar nucleon mass. The subsequent air shower is steered by hadron-nucleus collisions with low momentum transfer in
the non-perturbative regime of QCD. The shower description is
therefore based on phenomenological models, which are inspired by
1/$N$ QCD expansion and are also supplemented with generally accepted
theoretical principles like duality, unitarity, Regge behavior, and
parton structure. But of course, these phenomenological models are
inherently uncertain due to: {\it (i)}~the lack of a fundamental theoretical
description of soft hadronic and nuclear interactions; {\it (ii)}~the
large extrapolation required from collider energies to the UHECR
range. For example, a major source of uncertainty in modeling UHECR interactions is encrypted in the extrapolation of
the measured parton densities several orders of magnitude down to low Bjorken-$x$. UHECRs that impact on the upper atmosphere with $E_N \sim 10^{9.7}~{\rm
  GeV}$ yield partons with $x \sim m_\pi/\sqrt{s_{NN}} \sim 10^{-6}$,
{\it viz.} far smaller than anything accessible at today's accelerators.

Along this line, data from the Pierre Auger Observatory suggest that the
hadronic component of showers (initiated by cosmic rays with energy
$10^{9.8} < E/{\rm GeV} < 10^{10.2}$) contains about 30\% to 60\% more
muons than expected from simulations based on hadronic interaction
models tuned to accommodate LHC data~\cite{PierreAuger:2016nfk}. The
significance of the discrepancy between data and model prediction is
somewhat above $2.1 \sigma$. The discrepancy between experiment and
simulations has also been observed in the Telescope Array
data analysis~\cite{TelescopeArray:2018eph}.\footnote{We note that the muon
  deficit of simulated events is not observed in
IceTop data with $10^{6.4} < E/{\rm GeV} <
10^{8.1}$~\cite{IceCube:2022yap}. However, the c.m. energy per nucleon
pair of these showers is well
below those of LHC collisions.}

To date the best hope for resolving the muon mismatch involves decreasing the energy fraction lost to photon production (dominantly
from $\pi^0$ decay) in the description of hadronic collisions, because this is the only
known modification of the cascade development which increases the mean muon number without
encountering any restrictions from other shower
observables~\cite{Allen:2013hfa}. The required suppression of pion
production could come from an enhancement of strangeness
production~\cite{Anchordoqui:2016oxy}.\footnote{A point worth noting
  at this juncture is that the strange fireball model put
  forward in~\cite{Anchordoqui:2016oxy} could only be the carrier of subdominant
   effects~\cite{Anchordoqui:2019laz}.}  Indeed, one could expect heavy flavor
production to be enhanced in kinematic regimes where quark masses may be insignificant.

The amount of forward strangeness production is traced by the ratio of
charged kaons to pions, for which the ratio of electron and muon
neutrino fluxes is a proxy that will be measured by FASER$\nu$2 and FLArE. Note that 
pions primarily decay into muon neutrinos, whereas kaon decays lead to
fluxes of both electron and muon neutrinos. On top of that, muon and
electron neutrinos with different parent mesons populate
different energy regions, and so the spectral shape can be used to
disentangle the neutrino origin. Moreover, given that
$m_\pi < m_K$, neutrinos from pion decay are more concentrated
around the line-of-sight than those of kaon origin, and hence
neutrinos from pions obtain less additional
transverse momentum than those from kaon decays. Accordingly, the
closeness of the neutrinos to the line-of-sight, or equivalently their
rapidity distribution, can also be used to disentangle different neutrino
origins and estimate the pion to kaon ratio.

\begin{figure*}
  \postscript{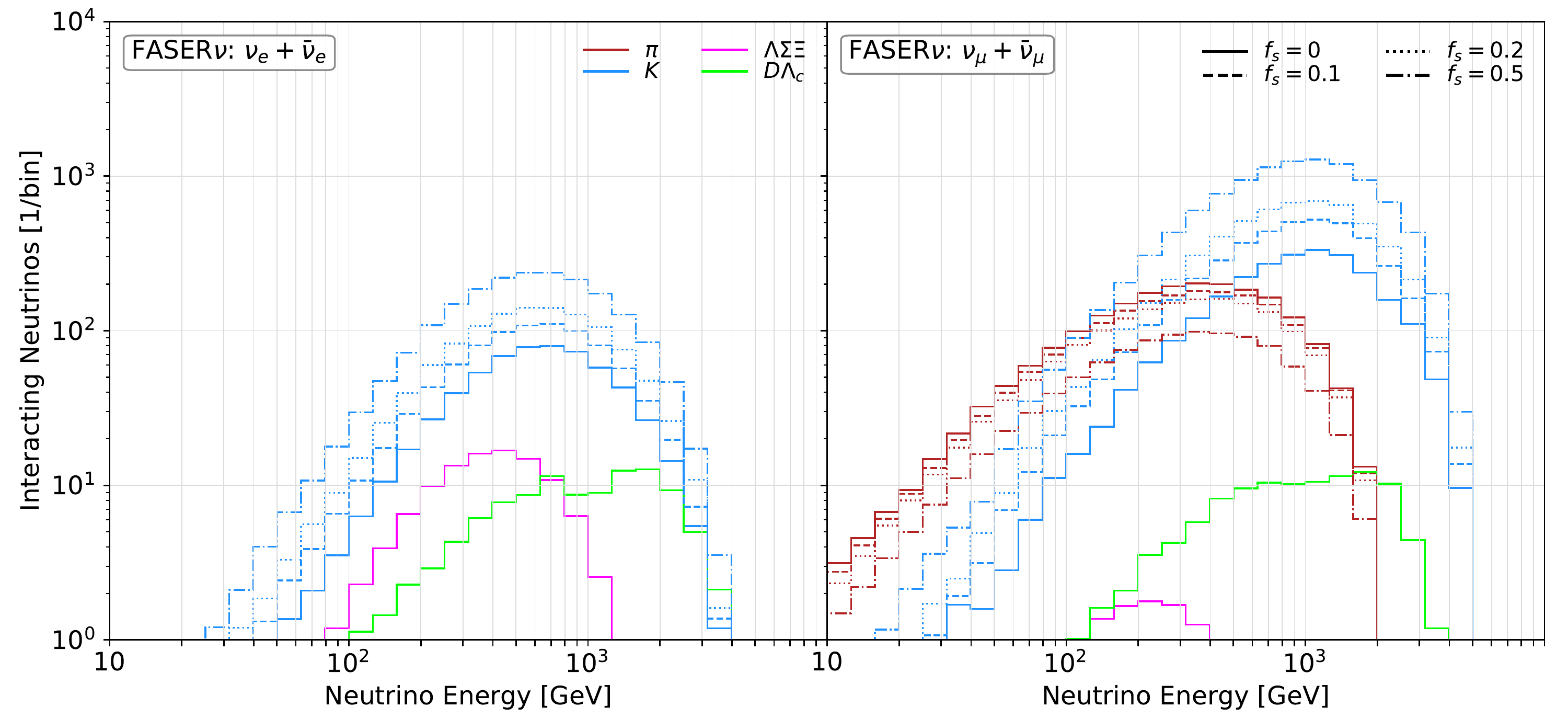}{0.9}
  \postscript{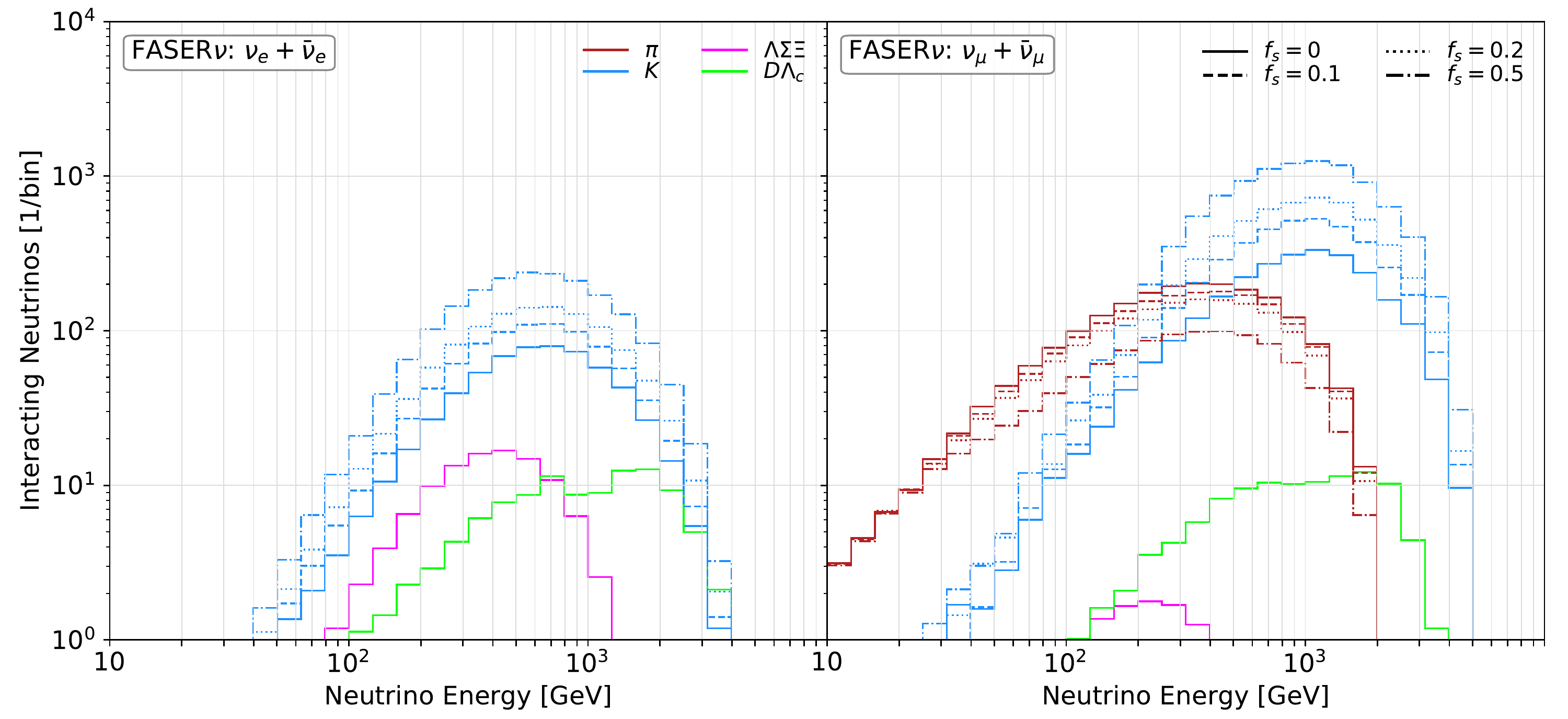}{0.9}
  \postscript{Figure_FASERv_eta8}{0.9}
  \caption{Energy spectrum of electron neutrinos (left) and muon neutrinos 
(right) interacting with FASER$\nu$. The vertical axis shows
the number of charged current neutrino interactions per energy 
bin for an integrated luminosity of $150~{\rm fb}^{-1}$
by different colors: pion decays (red), kaon decays (blue), hyperon
decays (magenta), and charm decays (green). The different line styles
correspond to predictions obtained from {\tt Sibyll-2.3d} with secondary pions
processed using  $F_s(\eta)$, for different values of $f_s$. We have taken $\eta_0 =
4,6,8$ from top to bottom. The image on the 1st row is taken
from~\cite{Anchordoqui:2022fpn}, while the images on the 2nd and 3rd rows are courtesy of Felix Kling.
 \label{fig:4}}
  \end{figure*}

\begin{figure*}
  \postscript{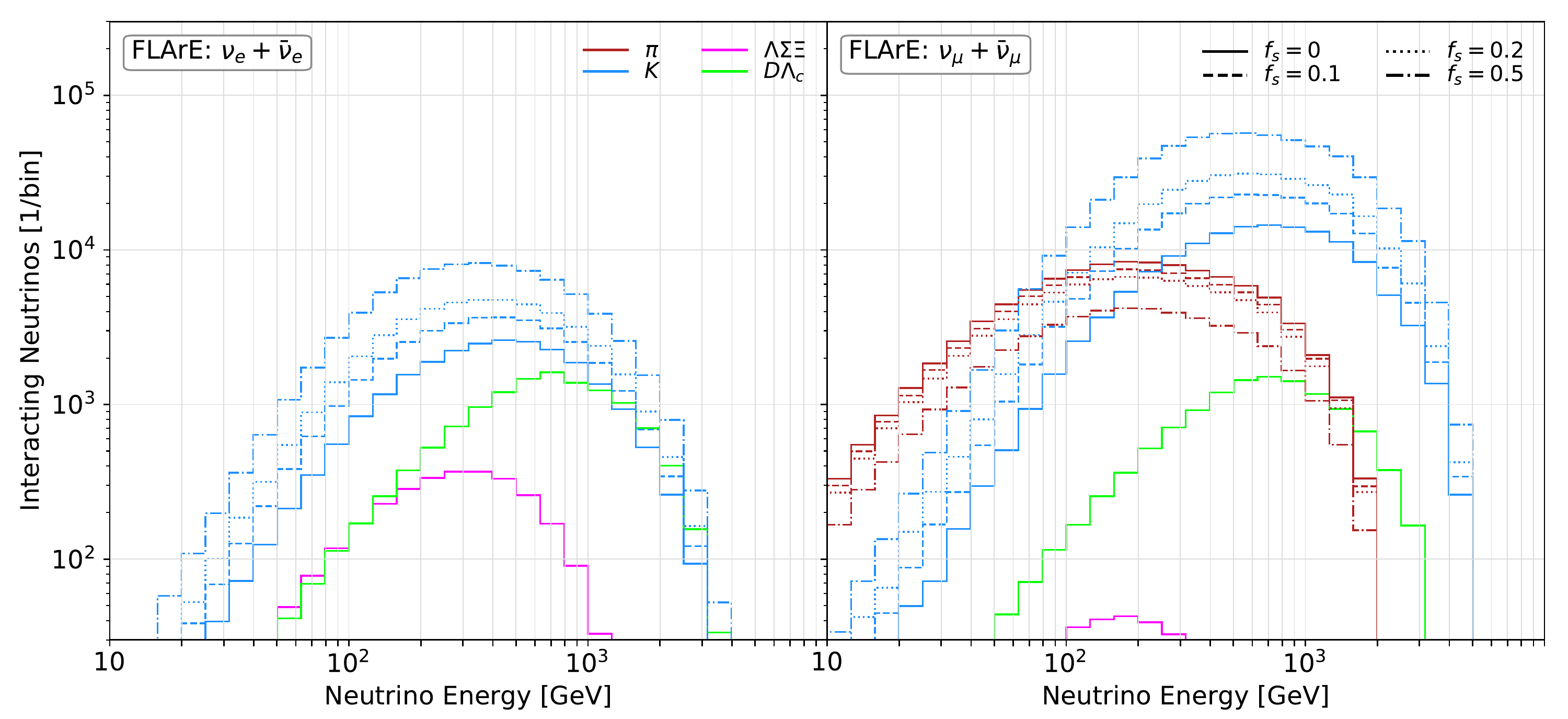}{0.9}
  \postscript{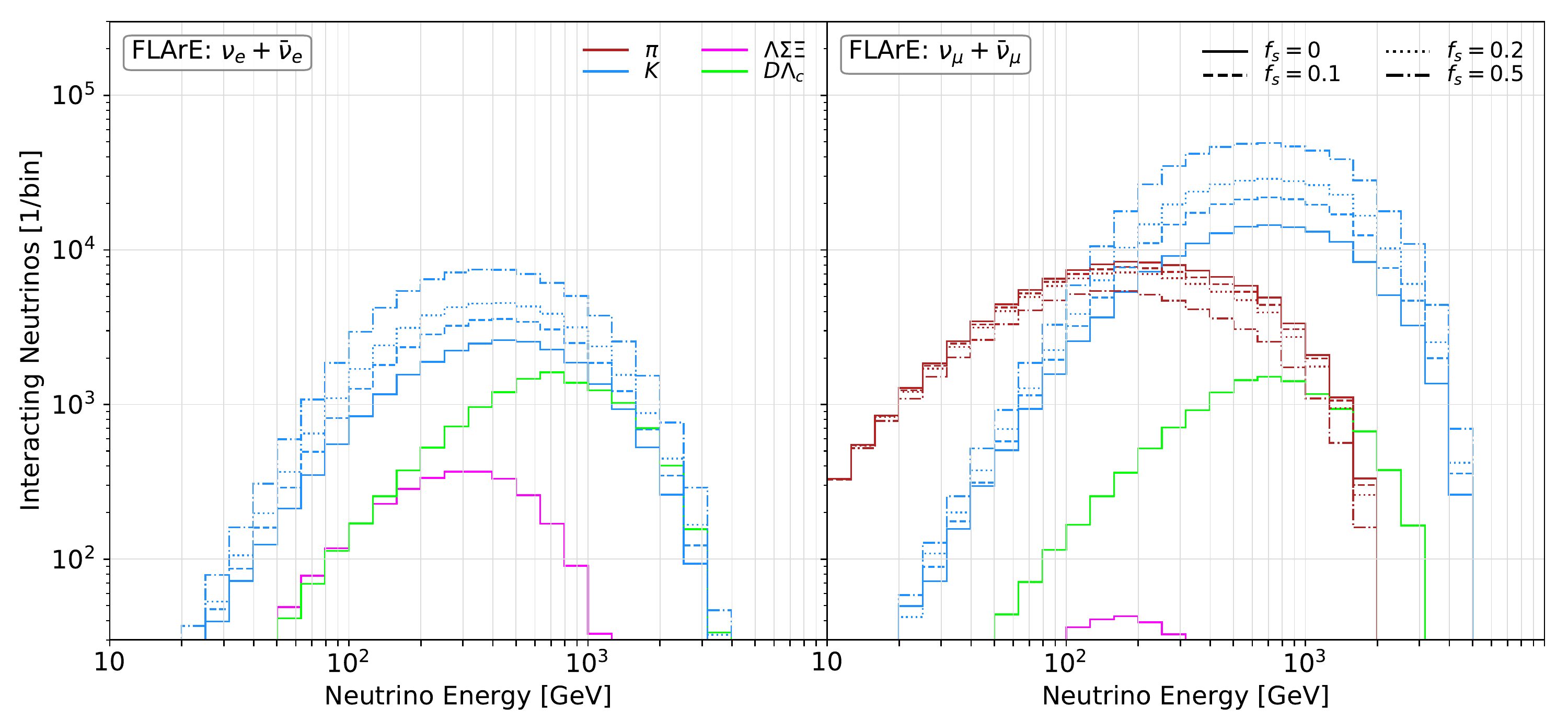}{0.9}
  \postscript{Figure_FLArE10_eta8}{0.9}
\caption{Expected number of charged current neutrino interactions
with the FLArE detector at the FPF assuming an integrated luminosity 
of $3~{\rm ab}^{-1}$. See Fig.~\ref{fig:4} for details.
\label{fig:5}}
  \end{figure*}

With this in mind, we estimate the sensitivity of the FPF experiments for measuring forward
strangeness production by looking at  the  pion-to-kaon ratio. In Fig.~\ref{fig:4}, we show the expected number of neutrino 
interactions with the FASER$\nu$ detector, assuming a
25~cm~$\times$~25~cm cross sectional area and a 1.2~ton target 
mass, as a function of the neutrino energy. Here, we have 
used {\tt Sibyll-2.3d}~\cite{Riehn:2019jet} as primary generator and 
use the fast LHC neutrino flux simulation 
introduced in~\cite{Kling:2021gos} to describe the propagation 
and decay of long-lived hadrons in the LHC beam pipe.\footnote{The
  hadronic interaction model {\tt
    Sibyll-2.3d} yields the smallest muon mismatch between data and simulations~\cite{Sciutto:2019pqs}.} The origin of
the neutrinos is indicated by the different line colors: red for 
pion decay, blue for kaon decay, magenta for hyperon decay, 
and green for charm decay. As explained above, the neutrinos from pions
and kaons populate different regions of phase space, which can be
used to disentangle pion and kaon production. In Fig.~\ref{fig:5}, 
we show the results for the FLArE detector at the FPF, which 
is assumed to have a 1~m$~\times$~1~m cross sectional area and a 10~ton 
target mass.

\begin{figure}[tpb]
  \postscript{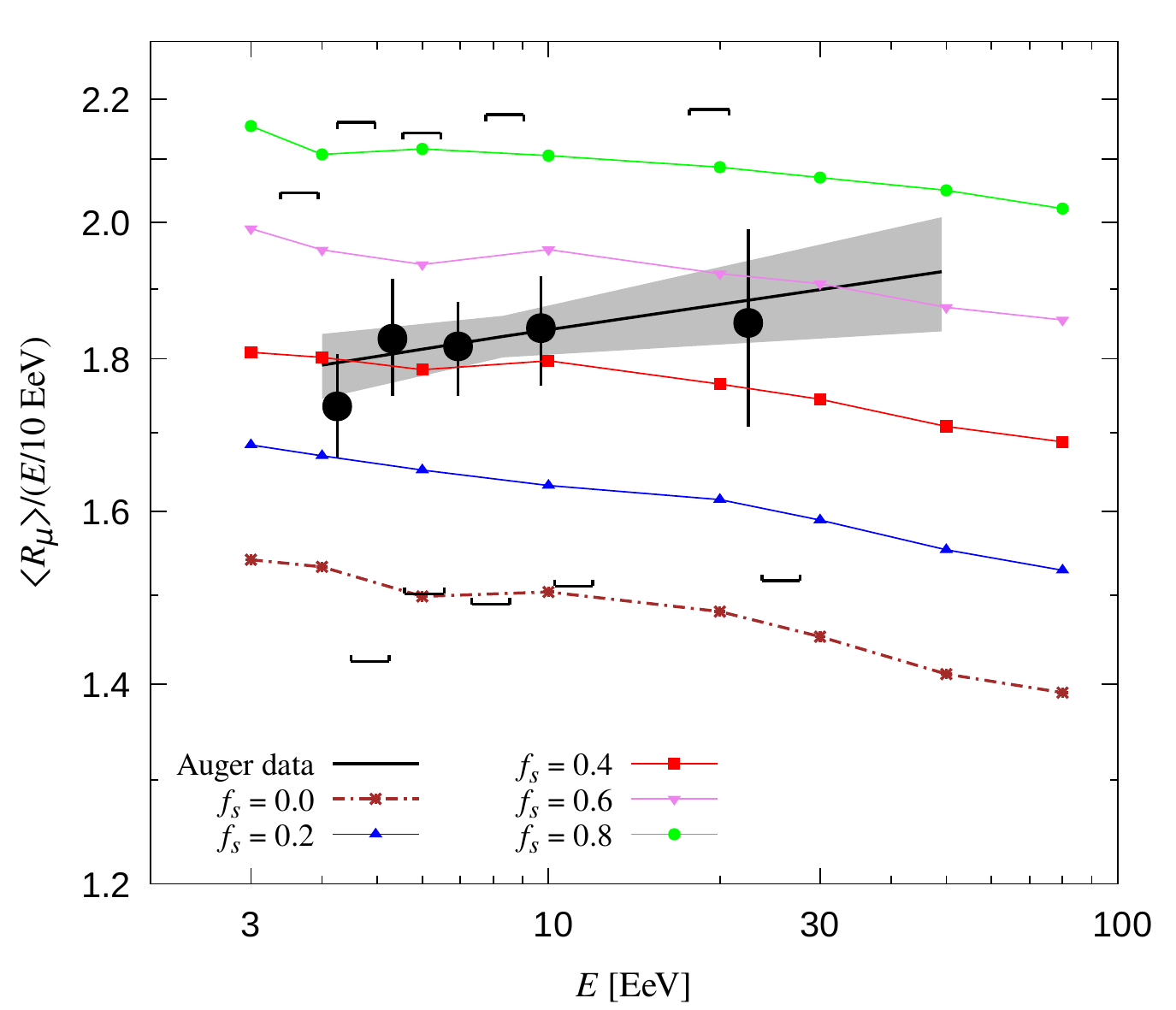}{0.5}
  \caption{The dimensionless muon content $\langle R_\mu \rangle = N_\mu/N_{\mu,10}$, where $N_\mu$ is the total number of muons 
    at ground level and $N_{\mu,10}$ is the
    average number of muons in simulated showers of protons with
    $E = 10^{10}~{\rm  GeV}$ and incident angle of
    $67^\circ$. $\langle R_\mu \rangle$ results from air shower simulations
for different values of 
  $f_s$ and $\eta_0 = 4$ are superimposed over Auger data with statistical
  \mbox{($\hspace{0.1em}\bullet\hspace*{-.66em}\mid\hspace{0.16em}$)}
  and systematic (\protect\rotatebox{90}{\hspace{-.075cm}[ ]})
  uncertainties~\cite{Aab:2014pza}. The simulations have been carried
  out with {\tt AIRES}~\cite{Sciutto:1999jh} {\tt + Sibyll-2.3d}
  assuming a nuclear composition of UHECRs derived from partial fluxes of
  the fit reported by the Pierre Auger
  Collaboration~\cite{PierreAuger:2016use}. For deatils,
  see~\cite{Anchordoqui:2022fpn}. \label{fig:6}}
\end{figure}

In Fig.~\ref{fig:4} and Fig.~\ref{fig:5}, we also show how a 
$\pi\leftrightarrow K$ swapping with probability given by
\begin{equation}
  F_s (\eta) = \left\{ \begin{array}{ccc}
        f_s & {\rm if} & -\infty < \eta < -\eta_0 \\
                        0 & {\rm if} & -\eta_0 \leq \eta \leq \eta_0 \\
                        f_s & {\rm if} & \phantom{-} \eta_0 < \eta < \infty
\end{array} \right. \,,                        \end{equation}
changes
the expected neutrino fluxes and event rates for the considered
experiments, where $\eta$ is the pseudorapidity in the center-of-mass frame,
$\eta_0 = 4, 6, 8$ and $0 < f_s < 1$. For $f_s >0$,
the particle species are
  changed according to the following criteria:
{\it (i)}~each $\pi^0$ is transformed onto $K^0_S$ or $K^0_L$, with
50\% chance between them; {\it (ii)}~each $\pi^+$ ($\pi^-$) is transformed onto $K^+$ ($K^-$)~\cite{Anchordoqui:2022fpn}. As expected,  positive values of $f_s$ lead to a 
suppression of the neutrino flux from pions as well as a larger 
relative enhancement of the neutrino flux from kaons. This is due 
to the  initially roughly 10 times larger flux of pions, 
such that even a small rate of $\pi \leftrightarrow K$ swapping 
can substantially increase the neutrino flux from the kaon decays. 
This leads to the remarkable result that already for $f_s=0.1$ 
($f_s=0.2$) the predicted electron neutrino flux at the peak of the 
spectrum is a factor of 1.6 (2.2) larger. These differences are 
significantly larger than the anticipated statistical uncertainties 
at the FPF~\cite{Kling:2021gos, Anchordoqui:2021ghd}.
We can see that there is almost no variation in the neutrino flux
while changing $\eta_0$ from
4 to 6. Requiring $\eta_0 = 8$ mainly affects the lower energy part of
the neutrino spectrum, and the effect of non-zero $f_s$ becomes
somewhat 
smaller. Altogether, we can conclude that FASER$\nu$ and FPF neutrino flux measurements will provide invaluable information 
to measure the rate of forward strangeness production through a measurement of
the pion-to-kaon ratio. This measurement could become a unique asset
in addressing the UHECR muon conundrum; see Fig.~\ref{fig:6}.

\subsection{Forward charm production}

\begin{figure}
  \postscript{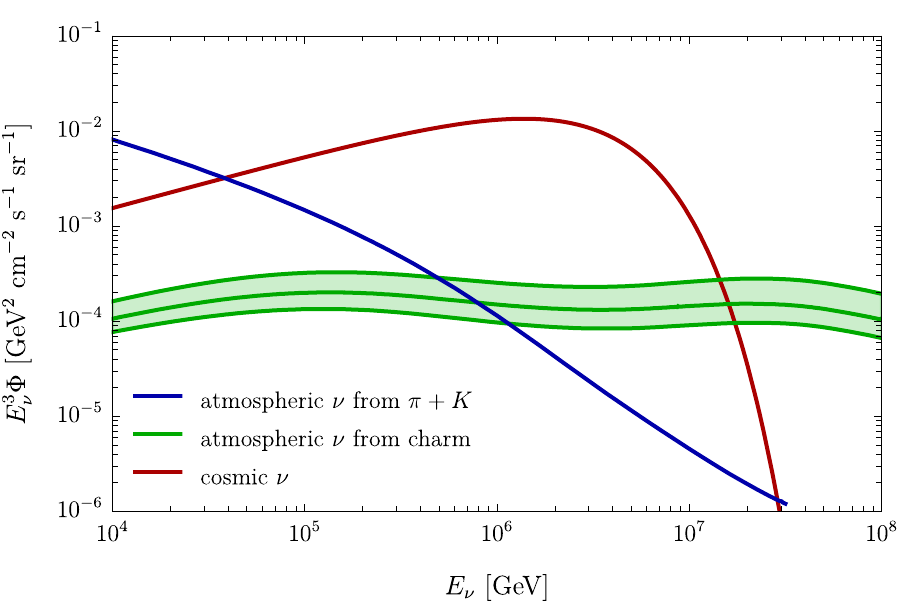}{0.5}
  \caption{Atmospheric and cosmic neutrinos fluxes per neutrino flavor
    scaled by $E_\nu^3$. The blue curve shows the atmospheric neutrino
    flux produced via decay of pions and kaons~\cite{Gaisser:2002jj}. The green band indicates the prompt
    atmospheric neutrino flux from perturbative charm production and
    its uncertainty~\cite{Bhattacharya:2015jpa}. The red curve denotes the flux of cosmic neutrinos
    from a fit to 6~yr of IceCube data~\cite{IceCube:2020acn}. IceCube's 7.5~yr sample gives
very similar results~\cite{IceCube:2020wum}. \label{fig:7}}
\end{figure}

Decays of mesons produced in air showers lead to a large flux of
atmospheric neutrinos. The decay of charged pions determines the neutrino
energy spectra up to
about 100~GeV above which they become increasingly modified by the kaon contribution,
which asymptotically reaches 90\%. As can be seen in Fig.~\ref{fig:7}, above about $10^{5.5}~{\rm GeV}$, kaons
are also significantly attenuated before decaying and the ``prompt''
component, arising mainly from very short-lived charmed hadrons ($D^\pm$,
 $D^0$, $\bar D^0$, $D_s^\pm$, $B_c^+$, and $\Lambda_c$)
dominates the spectrum~\cite{Anchordoqui:2005is}. The flavor ratio of these prompt neutrinos is $N_{\nu_e} :
N_{\nu_\mu} : N_{\nu_\tau} \sim 12/25 : 12/25 : 1/25$.

The neutrino flux arising from pion and kaon decay is reasonable well
understood, with an uncertainty in the range
10\% - 20\% below $10^{5.5}~{\rm GeV}$~\cite{Gaisser:2002jj}. However, the understanding of the
prompt neutrino flux is hampered by the large uncertainties in the QCD
modeling of heavy meson production~\cite{Halzen:2016thi}. The cosmic neutrino flux is also
subject to large uncertainties. A fit to IceCube cascade data assuming a single power law spectrum with a cutoff yields,
\begin{equation}
  \Phi = 3 \times 10^{-18} \ \Phi_0 \
  \left(\frac{E_\nu}{E_0}\right)^{-\gamma}  \ e^
  {-E_\nu/E_{\rm cut}}~({\rm GeV \, cm^2 \, s \, sr})^{-1},
\end{equation}  
where $E_\nu$ is the neutrino energy, $\Phi_0 = 1.83^{+0.37}_{-0.31}$,
$\gamma = 2.45^{+ 0.09}_{-0.11}$, $\log_{10} (E_{\rm cut}/{\rm GeV}) =
6.4^{+0.9}_{-0.4}$, and
$E_0 = 100~{\rm TeV}$~\cite{IceCube:2020acn}. Note that the
prompt neutrino flux could dominate the cosmic neutrino flux 
above $10^{7.3}~{\rm GeV}$. Besides, a major feature of the UHECR spectrum above
$10^{10}~{\rm GeV}$ is the increasing fraction of heavy nuclei with
respect to light nuclei~\cite{Watson:2021rfb}. This implies that the
cosmogenic neutrino flux could be strongly
suppressed~\cite{Anchordoqui:2007fi} and we may be witnessing the
high-energy cutoff of cosmic neutrinos in IceCube data  (without the need for Lorentz invariance violation~\cite{Anchordoqui:2014hua}). All in all, it seems critical to
obtain a robust measurement of the atmospheric charm
production.

A determination of the prompt neutrino flux requires a calculation of
the charm production differential cross-section $d\sigma/dx_c$ followed by the
hadronization of charm particles, where $x_c$ is the longitudinal
energy fraction of the produced charm quark~\cite{Bhattacharya:2015jpa}. Alternatively one uses
the Feynman-$x$ variable
\begin{equation}
  x_c \simeq x_F \equiv \frac{p_L^*}{|{\rm max} (p_L^*)|} \simeq \frac{2
    p_L^*}{\sqrt{s_{NN}}} = \frac{2 m_T \ \sinh y^*}{\sqrt{s_{NN}}}
  \label{xF}
\end{equation}
where $m_T \sim 3~{\rm GeV}$ is the transverse mass,
 and
$p^*_L$ and $y^*$ are the longitudinal momentum and 
rapidity in the center-of-mass frame. Now, {\tt Sibyll-2.3c} predicts that for  $E_\nu \sim
10^7~{\rm GeV}$, the probability density function of the parent nucleon energies peaks
at \mbox{$E_N \sim 30 \, E_\nu$~\cite{Fedynitch:2018cbl}.} If we  approximate the
energy of the charm meson as $E_c \sim 3 E_\nu$, it follows that $x_c \equiv E_c/E_N \sim 0.1$. For large
rapidity, $\sinh y^* \simeq  e^{y^*}/2$ and so
(\ref{xF}) leads to
\begin{equation}
  y^* \simeq \ln \left(\sqrt{6 \ m_N \ E_\nu \ x_c}/m_T \right) \, .
\label{y*}
\end{equation}
Using (\ref{y*}) it is straightforward to verify that neutrinos with $10^7 < E_\nu/{\rm
  GeV} < 10^8$ span the forward rapidity range $6.2 < y^* < 7.8$.
 
All told, LHC neutrinos to be measured at the FPF experiments could give critical
information on charm production at Feynman-$x$
close to 1. This process could potentially become a source of
background for cosmic neutrinos with $E_\nu \gtrsim 10^{7.3}~{\rm GeV}$, and at the moment we have no data and we have no theory for the process. 

\section{Conclusions}
The next breakthrough in particle physics is likely to involve
LLPs. FPF experiments operating at the HL-LHC will be sensitive to an
unexplored phase space for a broad range of LLP hidden sector physics.
In addition, neutrino measurements at these experiments will
improve the modeling of high-energy hadronic interactions in the
atmosphere and consequently will reduce the associated uncertainties of air shower measurements.

\section*{Acknowledgments}   
It is a pleasure to acknowledge many inspiring and fruitful
discussions with Ignatios Antoniadis, Karim Benakli, Jonathan
Feng, Carlos Garc\'{i}a Canal, Felix Kling, Dieter L\"ust, Hallsie
Reno, Sergio Sciutto, and Jorge Fernandez Soriano. This work has been supported by
the U.S. NSF PHY-2112527 and NASA 80NSSC18K0464.


\begin{thebibliography}{99}


\bibitem{Anchordoqui:2021ghd}
L.~A.~Anchordoqui \textit{et al.},
Phys. Rept. \textbf{968}, 1 (2022).


\bibitem{Feng:2022inv}
J.~L.~Feng \textit{et al.},
{arXiv:2203.05090}.

\bibitem{Jonathan:2019}
  J. L. Feng, talk at {\it Cosmic Controversies}, Kavli Institute for
  Cosmological Physics, Chicago (2019).


\bibitem{Feng:2010gw}
J.~L.~Feng,
Ann. Rev. Astron. Astrophys. \textbf{48}, 495 (2010).


\bibitem{Feng:2017uoz}
J.~L.~Feng, I.~Galon, F.~Kling and S.~Trojanowski,
Phys. Rev. D \textbf{97}, 035001 (2018).







\bibitem{Jonathan:2018}
  J.~L.~Feng, talk at {\it New Probes for Physics Beyond the Standard
    Model}, KITP, Santa Barbara (2018).

\bibitem{Holdom:1985ag}
B.~Holdom,
Phys. Lett. B \textbf{166}, 196 (1986).

\bibitem{Ilten:2016tkc}
P.~Ilten, Y.~Soreq, J.~Thaler, M.~Williams and W.~Xue,
Phys. Rev. Lett. \textbf{116}, 251803 (2016)



\bibitem{Kling:2021fwx}
F.~Kling and S.~Trojanowski,
Phys. Rev. D \textbf{104}, 035012 (2021)


\bibitem{Antoniadis:2001sw}
I.~Antoniadis, S.~Dimopoulos and A.~Giveon,
JHEP \textbf{05}, 055 (2001)


\bibitem{Anchordoqui:2022kuw}
L.~A.~Anchordoqui, I.~Antoniadis, K.~Benakli and D.~L\"ust,
{arXiv:2204.06469}.


\bibitem{Patt:2006fw}
B.~Patt and F.~Wilczek,
{arXiv:hep-ph/0605188}.



\bibitem{Pospelov:2007mp}
M.~Pospelov, A.~Ritz and M.~B.~Voloshin,
Phys. Lett. B \textbf{662}, 53-61 (2008).

\bibitem{Nomura:2008ru}
Y.~Nomura and J.~Thaler,
Phys. Rev. D \textbf{79}, 075008 (2009).

\bibitem{Batell:2021blf}
B.~Batell, J.~L.~Feng and S.~Trojanowski,
Phys. Rev. D \textbf{103},  075023 (2021).



\bibitem{FASER:2019aik}
  A.~Ariga \textit{et al.},
{arXiv:1901.04468}.

\bibitem{Goldberg:1986nk}
H.~Goldberg and L.~J.~Hall,
Phys. Lett. B \textbf{174}, 151 (1986).

\bibitem{Foroughi-Abari:2020qar}
S.~Foroughi-Abari, F.~Kling and Y.~D.~Tsai,
Phys. Rev. D \textbf{104}, 035014 (2021).



\bibitem{Alekhin:2015byh}
S.~Alekhin \textit{et al.},
Rept. Prog. Phys. \textbf{79}, 124201 (2016).

\bibitem{Cerci:2021nlb}
S.~Cerci \textit{et al.},
{arXiv:2201.00019}.
  
\bibitem{MATHUSLA:2019qpy}
  H.~Lubatti \textit{et al.},
JINST \textbf{15}, C06026 (2020).

\bibitem{FASER:2021mtu}
  H.~Abreu \textit{et al.},
Phys. Rev. D \textbf{104}, L091101 (2021).


\bibitem{Anchordoqui:2018qom}
L.~A.~Anchordoqui,
Phys. Rept. \textbf{801}, 1 (2019).

\bibitem{PierreAuger:2016nfk}
  A.~Aab \textit{et al.},
Phys. Rev. Lett. \textbf{117}, 192001 (2016).

\bibitem{TelescopeArray:2018eph}
  R.~Abbasi \textit{et al.},
Phys. Rev. D \textbf{98}, 022002 (2018).

\bibitem{IceCube:2022yap}
  R.~Abbasi \textit{et al.},
{arXiv:2201.12635}.

\bibitem{Allen:2013hfa}
J.~Allen and G.~Farrar,
{arXiv:1307.7131}.

\bibitem{Anchordoqui:2016oxy}
L.~A.~Anchordoqui, H.~Goldberg and T.~J.~Weiler,
Phys. Rev. D \textbf{95}, 063005 (2017).

\bibitem{Anchordoqui:2019laz}
L.~A.~Anchordoqui, C.~Garc\'\i{}a Canal, S.~J.~Sciutto and J.~F.~Soriano,
Phys. Lett. B \textbf{810}, 135837 (2020).

\bibitem{Riehn:2019jet}
F.~Riehn, R.~Engel, A.~Fedynitch, T.~K.~Gaisser and T.~Stanev,
Phys. Rev. D \textbf{102}, 063002 (2020).


\bibitem{Kling:2021gos}
F.~Kling and L.~J.~Nevay,
Phys. Rev. D \textbf{104}, 113008 (2021).

\bibitem{Sciutto:2019pqs}
S.~J.~Sciutto,
EPJ Web Conf. \textbf{210}, 02007 (2019).



\bibitem{Anchordoqui:2022fpn}
L.~A.~Anchordoqui, C.~Garc\'{\i}a Canal, F.~Kling, S.~J.~Sciutto and J.~F.~Soriano,
JHEAp \textbf{34}, 19 (2022).


\bibitem{Aab:2014pza} 
  A.~Aab {\it et al.},
  Phys.\ Rev.\ D {\bf 91}, 032003 (2015)
  [Phys.\ Rev.\ D {\bf 91}, 059901 (2015)].

\bibitem{Sciutto:1999jh}
S.~J.~Sciutto,
{arXiv:astro-ph/9911331};\\
{\tt http://aires.fisica.unlp.edu.ar}.
 
\bibitem{PierreAuger:2016use}
  A.~Aab \textit{et al.},
JCAP \textbf{04}, 038 (2017)
[JCAP \textbf{03}, E02 (2018)].


\bibitem{Gaisser:2002jj}
T.~K.~Gaisser and M.~Honda,
Ann. Rev. Nucl. Part. Sci. \textbf{52}, 153 (2002).


\bibitem{Bhattacharya:2015jpa}
A.~Bhattacharya, R.~Enberg, M.~H.~Reno, I.~Sarcevic and A.~Stasto,
JHEP \textbf{06}, 110 (2015).


\bibitem{IceCube:2020acn}
  M.~G.~Aartsen \textit{et al.},
Phys. Rev. Lett. \textbf{125}, 121104 (2020).

\bibitem{IceCube:2020wum}
  R.~Abbasi \textit{et al.},
Phys. Rev. D \textbf{104}, 022002 (2021).

\bibitem{Anchordoqui:2005is}
L.~Anchordoqui and F.~Halzen,
Annals Phys. \textbf{321}, 2660 (2006)


\bibitem{Halzen:2016thi}
F.~Halzen and L.~Wille,
Phys. Rev. D \textbf{94}, 014014 (2016).

\bibitem{Watson:2021rfb}
A.~A.~Watson,
JHEAp \textbf{33}, 14 (2022).

\bibitem{Anchordoqui:2007fi}
L.~A.~Anchordoqui, H.~Goldberg, D.~Hooper, S.~Sarkar and A.~M.~Taylor,
Phys. Rev. D \textbf{76}, 123008 (2007).



\bibitem{Anchordoqui:2014hua}
L.~A.~Anchordoqui, V.~Barger, H.~Goldberg, J.~G.~Learned, D.~Marfatia, S.~Pakvasa, T.~C.~Paul and T.~J.~Weiler,
Phys. Lett. B \textbf{739}, 99 (2014).

\bibitem{Fedynitch:2018cbl}
A.~Fedynitch, F.~Riehn, R.~Engel, T.~K.~Gaisser and T.~Stanev,
Phys. Rev. D \textbf{100}, 103018 (2019).


\end{thebibliography}
\end{document}